# Comparison of Self-monitoring Feedback Data from Electronic Food and Nutrition Tracking Tools

Ahmed Fadhil[a]

[a]*University of Trento*

**Abstract**

Changing dietary habits and keeping food diary encourages fewer calorie consumption, and thus weight loss. Studies have shown that people who keep food diary are more successful in losing weight and keeping it off. However, no study has investigated the nutritional values produced by food journaling applications. This is crucial since keeping food diaries helps identify areas where changes needed to help user's loss weight, based on the application feedback. To achieve this, the provided data should be consistent among all applications. Otherwise, this will question the effectiveness and reliability of such tools in tracking diet and weight loss, and hence question user trust in these applications.

This study characterises the use of 4 food journaling applications to track user diet for 10 days (namely, MyFitnessPal, Lose It!, FatSecret, CRON-O-Meter). We measured variations between the output of each application. The findings revealed an inconsistent and a variation in the output feedback given by all the 4 tools. Although some tools provided closer values, still their data were different and inconsistent. Moreover, some tools were missing essential nutritional fact data, such as sugar and fiber. We additionally compared a sample of food items common among all tools with the Swiss Food Composition Database and checked for their consistency with the same items in the database. The evaluation of the applications showed a gap in the data consistency among applications and the FCD, and questions how reliable they're for food logging and diet tracking. This study contributes to current research in health and wellbeing and can be referenced by researchers to provide deeper insights into the data consistency. Future work should examine ways to provide precise output that is common among all applications, so to guarantee the effect on weight loss.

## 1. Introduction

A strong correlation exists between dietary habits and intermediate health risk factors, such as diabetes, blood pressure and cholesterol [1]. Diet plays a crucial role in the development and prevention of cardiovascular disease; it's a key changing element that impacts

*Email address:* `ahmed.fadhil@unitn.it` (Ahmed Fadhil)

all healthy risk factors. Therefore, to achieve positive impact, a balanced and mindful approach of overall dietary pattern is required. Keeping daily food diary to track diet, lose weight or build healthy habits with a paper or an app and other forms of record have shown to positively impact weight lost, diet management, controlling portion size and sticking to healthy habits [2]. The act of meal recording helps think about the meal, the time, and the portion size [3]. A study by Hollis et al., [4] concluded that people who kept a food diary were more likely to keep the weight they lost off, in another study by Burke et al., [5] found that people who kept a record of their meals and kept up with diet and exercise lost nearly twice as much as people who didn't keep a log (but did diet and exercise). A combined emphasis on dietary intake and physical activity is important to short- and long-term weight loss goals.

There exists abundant list of mobile and web apps intended for daily diary creating and food journaling. These tools provide insight about caloric intake, the amount of nutritional values, and other dietary related data. The majority of these tools ask users to log their dietary intake, including item name, quantity and meal time. These applications are still tedious in terms of data entry and provide limited automation to detect and fill-in the data. In addition, the main issue with such tools is the nutritional facts output data. Such data have to be precise and consistent among all applications. For example, the caloric and amount of ingredient should be consistent along all the applications.

In this research we describe data variation and output of eating quantity by each tool. We focus on the amount of caloric data provided by each tool (namely, MyFitnessPal[1], Lose It![2], FatSecret[3] and CRON-O-Meter[4]) together with other nutritional values (namely, protein, sugar, fiber, fat, carbs) and compare them to the Swiss food composition database[5], then compare the final result in terms of portion size. We highlight data consistency given by each food journaling application. We begin by comparing their values and check them in terms of food weight and portion size. For that, we conducted an experimentation with 6 users and the 4 food journaling tools to track their food logging for 10 days. The user demographic consisted of 5 male and 1 female, all attending university level study. The findings revealed variation in output among all tools and with the Swiss database. We tracked daily meal, including breakfast, lunch, dinner and snacks. After data collection, we analysed the data by two reviewers. Missing or inconsistent values in one or more application were eliminated from the study. We analysed associated food list and nutrient values and compared them in terms of size and value similarity. Finally, a random list of 13 food items were selected and compared with the Swiss database in terms of portion size. There was a variation in the value provided by the tools and the database with respect to portion size.

Although the importance of diet tracking for overall health and wellness, to our knowledge no study has examined and validated the output provided by food journaling application and compared their various nutritional values. We envisage this study to contribute to the existing body of research in the domain of diet management, weight loss and preventive healthcare.

---

[1] http://www.MyFitnessPal.com/
[2] https://www.Lose It.com/
[3] https://www.FatSecret.it/
[4] https://CRON-O-Meter.com/
[5] http://naehrwertdaten.ch/

## 2. Research Study

*2.1. Electronic Diet Tracking*

Self-monitoring via mobile apps is a recent addition to self-regulation methods for weight loss. Self-monitoring exercise proven relevant to weight loss [6]. A study by Tsai et al., [6] describes Patient-Centered Assessment and Counselling Mobile Energy Balance (PmEB) application to self-monitor caloric balance in real time. The study concluded that PmEB is feasible for self-monitoring and weight management. A study by Noronha et al., [7] introduced PlateMate, an application that allows users take picture of their meal and receive estimates of nutritional facts. The study pointed the importance of data accuracy provided by such tools to help monitor dietary goals. However, current methods through self-reporting, expert observation or algorithmic analysis are time-consuming. Results showed that PlateMate is nearly as accurate as dietitian and easier to use by users. Yet, the data provided by mobile application often doesn't adhere to evidence-based content. Breton et al., [8] summarised the content of available weight control apps. Information on content, user ratings, and price were extracted. The findings showed many apps lack evidence-based content and suggested further research work.

Tracking diet or physical activity is a social process and requires collaborative process rather than personal. Different tracking styles exist, including goal driven and documentary tracking. Rooksby et al., [9] characterised activity trackers as "lived informatics". This is contrasting with other discussions on personal informatics and quantified-self. The study interviewed users who track activity and found people interweave various activity trackers, sometimes with the same functionality. Physical activity and diet monitoring are key for weight loss programs. The purpose of a study by Turner et al., [10] was to assess the relationship between diet (mobile app, website, or paper journal) and physical activity (mobile app vs no mobile app) monitoring. The findings pointed to the potential benefits of mobile monitoring methods during the weight loss trials. The study suggested ways to predict which self-monitoring method works better [10]. Current data generated by apps are used as incentives to improve health and lower associated cost. To help improve patient's health and fitness, data can be collected from smartphone's built-in tools, such as Global Positioning System (GPS) tracking, accelerometer, and pedometer to measure health and fitness parameters. These apps then analyse the data and summarize it, provide frequent feedbacks as well as devise individualised plans based on user goals [11]. Health telematics is a growing up issue that is becoming a major improvement on patient lives, especially in elderly. M-Health solutions address emerging problems on health services, including, the increasing number of chronic diseases related to lifestyle, high costs of care. Silva et al., [12] presented a comprehensive review of the state on m-Health services and applications. Open and challenging issues on emerging m-Health solutions were proposed for further works. Mobile devices offer media rich and context aware features that are useful for eHealth applications. A paper by Liu et al., [13] examined apps from App Store and provided a status of mHealth apps and an analysis of related technology. The study serves as a reference point and guide for developers and practitioners interested in using iOS as a platform for m-health applications, particular from the technical point of view. mobile apps offer the potential to objectively monitor user's eating and activity behaviours and encourage healthier lifestyles. In this regard, BALANCE [14] is an app for long-term wellness management. The system automatically detects caloric expenditure via sensor data from a Mobile Sensing Platform unit. Diet tracking is linked to improved weight loss success. Mobile apps could allow for improved dietary adherence [15] and manage various aspects of diseases, such as diabetes through consistent self-monitoring blood glucose (SMBG) [18]. Moreover, food journaling is an effective way to regulate excessive food intake. However, manual food journaling is burdensome, and crowd-assisted food journaling has been explored to ease this process. Rabbi et al., [16] proposed an approach to label food images with only high performing labelers, since they provide good quality labels. The study provided a machine learning algorithm to automatically identify high performing crowd-labelers from a dataset of 3925 images collected over 5 months. Self-logging is a critical component to many health applications. However, sustaining the data logging is a tedious process. A work by Bentley et al., [17] demonstrated passive mobile notifications to increase food intake logging. Adding notifications increased the frequency of logging from 12% in a one-month, ten-user pilot study without reminders to 63% in the full 60-user study with reminders included.

We considered 4 food journaling and nutrition tracking tools. These tools were selected after measuring factors related to their data collection, and the possibility to download and access these data. Moreover, we paid attention to the caloric values returned by these apps. We checked for the sentiment analysis features, namely perceived positiveness, addictiveness, crashes and negativeness of the selected apps as provided by AppTrace[6]. Below we list and define each point.

- *Positive:* Reviews that love the app or a feature, or generally find it fun, great, perfect, epic, etc. People recommending others to buy/download it.

- *Addictiveness:* Reviews talking about an app being enjoyable, useful, addictive, entertaining, fun, easy to use. People saying how they keep using an app or can't stop playing it.

- *Crash:* Reviews talking about the app or a feature crashing, freezing, not loading or working properly or needing bug-fixing.

- *Negative:* Reviews finding the app terrible, boring, unusable etc. People talking about deleting, uninstalling or even hating the app or a version; recommendations to others to not buy / download / upgrade.

*2.1.1. MyFitnessPal*

This mobile & web application tracks diet and physical activity to determine optimal caloric intake and nutrition for the user. This tool uses gamification elements to motivate user engagement during food logging. MyFitnessPal provides motivation and community empowerment to promote healthy living. The app tracks user's breakfast, lunch dinner and snacks. It provides comprehensive nutritional values about each food item, in addition to providing total amount of calories per nutrient value consumed. The tool provides nutritional data about calories consumed, carbs, fat, protein, cholesterol, sodium, sugar and fiber contained in each item. In Table-1 we list the sentiment analysis features of MyFitnessPal together with their rankings as provided by AppTrace.

---

[6]apptrace.com/

| Features | Ranking |
|---|---|
| *Positive* | 59% |
| *Addictiveness* | 16% |
| *Crash* | 10% |
| *Negative* | 8% |

Table 1: MyFitnessPal's Sentiment Analysis Features by AppTrace.

*2.1.2. Lose It!*

This app allows users to take picture of food and identify the food to get caloric count and nutritional information in a single snap. This app uses goal-setting techniques to help users achieve their weight loss, exercise, water intake and other health related activities. The application provides peer competition or joining a public challenge towards a common goal. The weight loss community is to provide inspiration, guidance, support and challenges. The tool tracks user daily meal, including also snacks and exercises. In addition, it provides a plan to achieve optimal goal. We obtained data about fat, cholesterol, sodium, carbohydrates, and protein from this app. In Table-2 we list the sentiment analysis features of Lose It! together with their rankings as provided by AppTrace.

| Lose It! | Ranking |
|---|---|
| *Positive* | 69% |
| *Addictiveness* | 26% |
| *Crash* | 2% |
| *Negative* | 3% |

Table 2: Lose It!'s Sentiment Analysis Features by AppTrace.

*2.1.3. FatSecret*

This app provides the possibility to track user daily meal, including snacks. The app uses community empowerment and focuses on speeding up food logging process. It displays fat, carbs, protein and calories about each food item, it also provides nutritional facts related to serving size and amount. In Table-3 we list the sentiment analysis features of FatSecret together with their rankings as provided by AppTrace.

| Features | Ranking |
|---|---|
| *Positive* | 46% |
| *Addictiveness* | 14% |
| *Crash* | 16% |
| *Negative* | 15% |

Table 3: FatSecret's Sentiment Analysis Features by AppTrace.

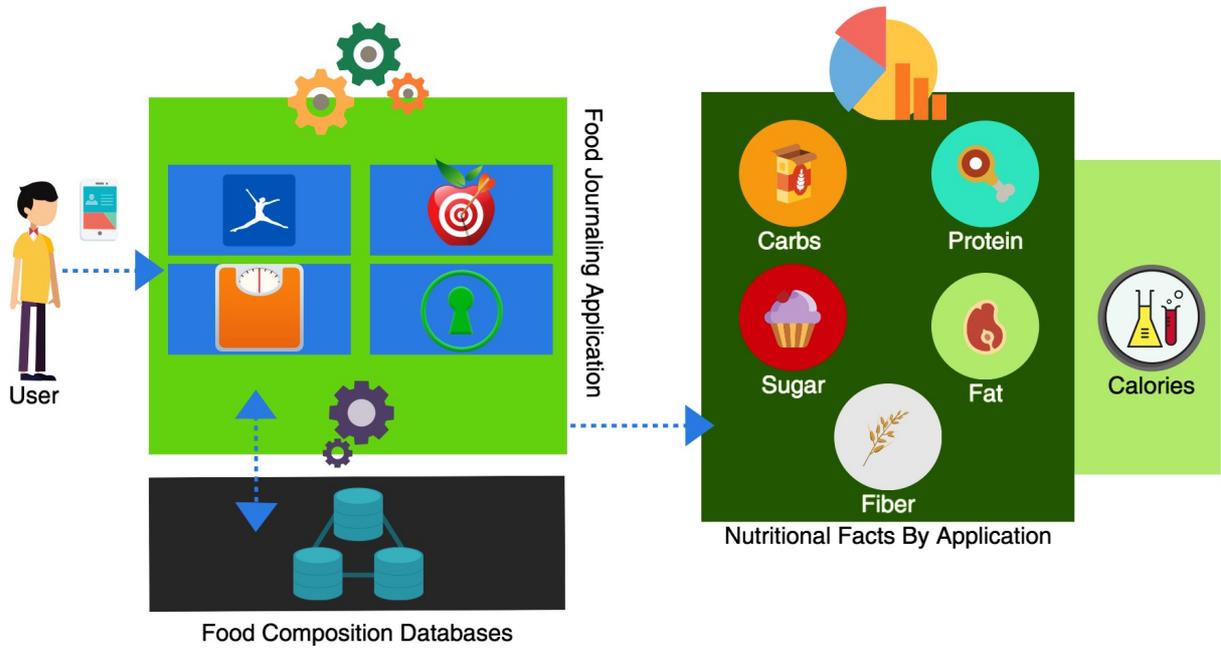

Figure 1: Food Journaling Application Analysis.

*2.1.4. CRON-O-Meter*

This app allows users to add food items, exercise and provides a summary of consumed calories and daily caloric limits based on user's BMI. The tool doesn't provide a way to insert periodic meals (e.g., breakfast, lunch, dinner). However, it provides great insights about 60+ micronutrients contained in each food item. This includes vitamins (e.g., Vitamin A, Vitamin B, Vitamin C), minerals (e.g., Calcium, Copper, Iron), lipids (e.g., Fat and Cholesterol), protein (e.g., Cystine, Histidine, Isoleucine) and carbs (e.g., Fiber and Sugar). The app synchs user nutritional data and activity with Apple's Health app. There was no sentiment analysis feature data about CRON-O-Meter. In Figure-1 we provide a pipeline overview for the process of user-application interaction to acquire the nutritional data. In Table-4 we highlight the food journaling tools considered in this study.

## 3. Research Analysis

This consists of the analysis steps undertaken to gather the data manually by users and save them in a file, then insert the data into each of the 4 food journaling applications.

| Food Journaling Applications | | | | |
|---|---|---|---|---|
| *Application* | *Ratings* | *Price* | *Category* | *Description* |
| *MyFitnessPal* | 5 - 5 | Free | Health & Fitness | Whether you want to lose weight, tone up, get healthy, change your habits, or start a new diet |
| *Lose It!* | 4 - 5 | Free | Health & Fitness | weight loss program |
| *FatSecret* | 4 - 5 | Free | Health & Fitness | weight loss and dieting program |
| *CRON-O-Meter* | 4 - 5 | e3,49 | Health & Fitness | online diary for tracking your diet, exercise, and other health related information |

Table 4: The User-application Pipeline.

After obtaining the result from the tools, search the data within the Swiss Food Composition Database and compare their output with the output obtained from the tools. Finally, we perform an analysis on the data significance and portion size estimation (see Figure-2 for the detailed data analysis followed to obtain the final result).

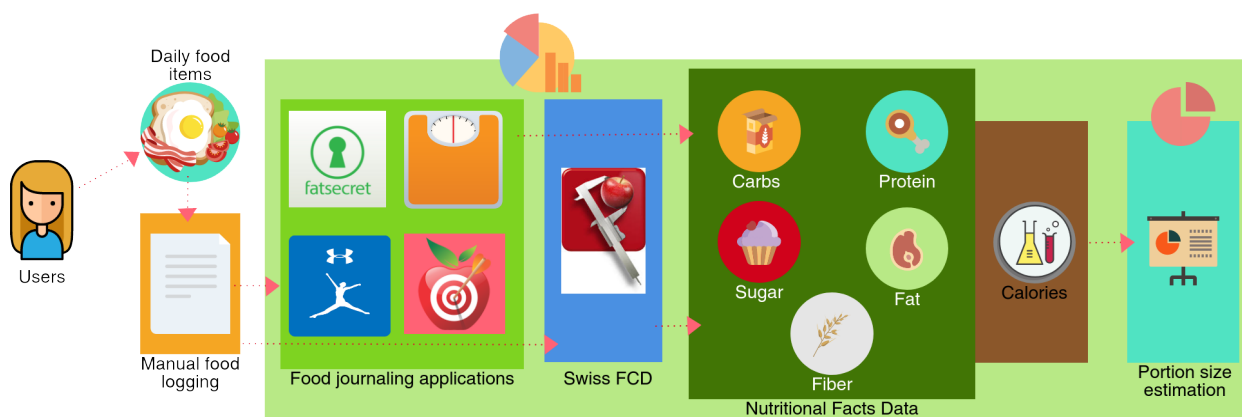

Figure 2: The Data Collection and Analysis.

### 3.1. Manual Data Collection

This step consisted of manually recording the dietary intakes by users for 10 days. The recording included brunch and snacks data during the day. The data was recorded in an Excel file by each user and included also data collected during the weekend. User's data was anonymized, however we kept track of their demographic information. Next, the data was prepared to be inserted into each food journaling application.

### 3.2. Food Journaling Data Analysis

After obtaining the data, we began analysing them with the food journaling applications. The data collection included all daily meals, snacks and brunches consumed throughout the day. The analysis focused on gathering data about carbs, protein, sugar, fat and fiber. The same food item was inserted in all apps and checked for the above nutritional values. The data for the same food item were checked and recorded in a separate file. However, some food items didn't appear and were not provided by some tools. For example, FatSecret had no data about fiber and sugar contained in the food items. Therefore, such food items were omitted from the tables. Moreover, Lose It! had no caloric data given for all food item, hence it was omitted from the caloric table. We considered the above nutritional information because they're greatly associated with healthy lifestyle and chronic conditions. For example, abnormal blood lipid (fat) levels have a strong correlation with coronary artery disease, heart attack and coronary death [19]. While high blood pressure (hypertension) is a major risk factor for cardiovascular disease, and it is associated with high sodium consumption [19]. An attention was given to the consistency in details of the food items. If an item was available in less than two tools, it was eliminated from the study. That way, we obtained a data consistent among all tools and eliminated data bias and inconsistency.

Having all food items in Italian, the only available food composition database supporting also Italian food items was the Swiss food composition database. Hence, all food items were searched in the Swiss database and recorded together with their nutritional information to be used later in the data analysis and comparison.

### *3.3. Data Comparison*

This step consisted of studying patterns among the output of food journaling applications. This includes getting users data for 10 days, then inserting the data into the 4 food journaling applications. We compared the output data for all the 4 applications. For the sake of consistency, we present the data for 3 users and 3 days per user, which is enough to validate and prove our hypothesis. Below we discuss the data comparison for the users. We checked the output for carbs, protein, sugar, fat and fiber for each food item among all the tools

### *3.3.1. Overall Nutritional Output*
**User #1**

The data analysis revealed incompatibility in the output for all the applications. The amount of carbs output was different in all the applications (see Table-5). For example, the carbs outputted by MyFitnessPal and Lose It! are inconsistent and have different output value for the same item. The carbs in MyFitnessPal is *188 g*, whereas it's *116.2 g* in Lose It! for the same day. Moreover, the data output by MyFitnessPal and CRON-O-Meter are closer in terms of quantity. Finally, some food items found to be common among some applications than others.

By checking the amount of sugar in Table-6, we obtained the same inconsistency among the data produced. For example, the amount of sugar in MyFitnessPal is *34 g*, whereas its *29.8 g* in Lose It! for the same day. Moreover, the sugar value was missing from FatSecret. In Table-7, we see the protein amount variation among all tools and there is a clear inconsistency in the output that shows the variation in protein content for the same food item. For example, by looking at the day-1 data, we see *74 g, 54.6 g, 62.62 g*, and *68.5 g* calculated by MyFitnessPal, Lose It!, FatSecret and CRON-O-Meter, respectively. The same result was obtained about fat and fiber as seen in Table-8 and Table-9. Finally, the amount of fiber was not provided by FatSecret.

|                | Day 1    | Day 2     | Day 3     |
|----------------|----------|-----------|-----------|
| **MyFitnessPal** | 225 g    | 188 g     | 225 g     |
| **Lose It!**     | 140.9 g  | 116.2 g   | 225.8 g   |
| **FatSecret**    | 168.3 g  | 125.86 g  | 194.32 g  |
| **CRON-O-Meter** | 203.8 g  | 200.5 g   | 208.1 g   |

Table 5: The Carbs data comparison for User 1

|                | Day 1   | Day 2  | Day 3  |
|----------------|---------|--------|--------|
| **MyFitnessPal** | 34 g    | 31 g   | 47 g   |
| **Lose It!**     | 29.8 g  | 37 g   | 35.4 g |
| **FatSecret**    | -       | -      | -      |
| **CRON-O-Meter** | 63.6 g  | 55.4 g | 54.9 g |

Table 6: The Sugar data comparison for User 1

|                | Day 1   | Day 2   | Day 3   |
|----------------|---------|---------|---------|
| **MyFitnessPal** | 74 g    | 53 g    | 52 g    |
| **Lose It!**     | 54.6 g  | 66.2 g  | 65.5 g  |
| **FatSecret**    | 62.62 g | 56.78 g | 50.88 g |
| **CRON-O-Meter** | 68.5 g  | 88.1 g  | 48 g    |

Table 7: The Protein data comparison for User 1

|                | Day 1   | Day 2   | Day 3   |
|----------------|---------|---------|---------|
| **MyFitnessPal** | 41 g    | 33 g    | 23 g    |
| **Lose It!**     | 24.5 g  | 53.9 g  | 21.4 g  |
| **FatSecret**    | 47.7 g  | 34.83 g | 36.79 g |
| **CRON-O-Meter** | 21.6 g  | 54.2 g  | 24.3 g  |

Table 8: The Fat data comparison for User 1

|                | Day 1   | Day 2  | Day 3  |
|----------------|---------|--------|--------|
| **MyFitnessPal** | 25 g    | 11 g   | 15 g   |
| **Lose It!**     | 11 g    | 8.5 g  | 17.1 g |
| **FatSecret**    | -       | -      | -      |
| **CRON-O-Meter** | 15.6 g  | 15.4 g | 19.3 g |

Table 9: The Fiber data comparison for User 1

|              | Day 1   | Day 2   | Day 3    |
|--------------|---------|---------|----------|
| **MyFitnessPal** | *91 g*  | *156 g* | *142 g*  |
| **Lose It!**     | *53.3 g*| *168 g* | *97.4 g* |
| **FatSecret**    | *72.29 g*| *207.49 g* | *106.52 g* |
| **CRON-O-Meter** | *74.6 g*| *120 g* | -        |

Table 10: The Carbs data comparison for User 2

|              | Day 1   | Day 2   | Day 3   |
|--------------|---------|---------|---------|
| **MyFitnessPal** | *49 g*  | *19 g*  | *29 g*  |
| **Lose It!**     | *42.7 g*| *43.6 g*| *16.6 g*|
| **FatSecret**    | *28.13 g*| *35.8 g*| *17.74 g*|
| **CRON-O-Meter** | *34.6 g*| *19.4 g*| -       |

Table 12: The Protein data comparison for User 2

|              | Day 1  | Day 2   | Day 3   |
|--------------|--------|---------|---------|
| **MyFitnessPal** | *4 g*  | *10 g*  | *16 g*  |
| **Lose It!**     | *16 g* | *18.1 g*| *16.6 g*|
| **FatSecret**    | -      | -       | -       |
| **CRON-O-Meter** | *5.2 g*| *13.4 g*| *8.7 g* |

Table 11: The Sugar data comparison for User 2

|              | Day 1   | Day 2   | Day 3   |
|--------------|---------|---------|---------|
| **MyFitnessPal** | *18 g*  | *2 g*   | *10 g*  |
| **Lose It!**     | *5 g*   | *4.1 g* | *9.3 g* |
| **FatSecret**    | *6.50 g*| *9.12 g*| *6.14 g*|
| **CRON-O-Meter** | *16.2 g*| *3.1 g* | -       |

Table 13: The Fat data comparison for User 2

|              | Day 1  | Day 2  | Day 3  |
|--------------|--------|--------|--------|
| **MyFitnessPal** | *3 g*  | *12 g* | *4 g*  |
| **Lose It!**     | *6.4 g*| *9.2 g*| *5.4 g*|
| **FatSecret**    | -      | -      | -      |
| **CRON-O-Meter** | *6.1 g*| *7.4 g*| *3.7 g*|

Table 14: The Fiber data comparison for User 2

### User #2

The output data is similar to user #1. That said, food items have different outputs by each application. The amount of carbs (see Table-10) varies among all the tools. For example, the carb in Lose It! was *53.3 g*, whereas it was *72.29 g* in FatSecret obtained for the same day. Some values were not provided by some applications. For example, CRON-O-Meter had missing carbs value due to unavailability of the food item. Looking at Table-11, we notice differences in data scattered among all tools, however MyFitnessPal and Lose It! had outputs with closer values than CRON-O- Meter. Moreover, sugar values were missing from FatSecret.

The protein amount in Table-12 distributed over 3 days is inconsistent and varies among all tools. For example, the portion amount for day-1 was *42.7 g* and *28.13 g* by Lose It! and FatSecret, respectively. The same data inconsistency was found in the fat and fiber data, shown in Table-13 and Table-14. Finally, the amount of fiber was not provided by FatSecret.

### User #3

Similar to the above discussed data, there is also inconsistency in all output data for user #3. For example, looking at the value outputted by all tools, we see in the carbs outputted by Lose It! and FatSecret is *160.3 g* and *197.44 g*, respectively. In Table-16, the sugar value for Lose It! is *47.6 g* and is missing from FatSecret. In Table-17 and Table-18, the amount for Lose It! and FatSecret was *34.2 g* and *42.88 g* for the protein, and *19.1 g* and *30.54 g* for the fat obtained from the same food item. Whereas in Table-19 the amount of fiber for Lose It! was *11.5 g* and was missing in FatSecret.

|              | Day 1   | Day 2   | Day 3  |
|--------------|---------|---------|--------|
| MyFitnessPal | 286 g   | 112 g   | 69 g   |
| Lose It!     | 160.3 g | 120.3 g | 58.6 g |
| FatSecret    | 197.44 g| 109.6 g | 92.56 g|
| CRON-O-Meter | 140.6 g | 76.1 g  | 100.1 g|

Table 15: The Carbs data comparison for User 3

|              | Day 1   | Day 2  | Day 3  |
|--------------|---------|--------|--------|
| MyFitnessPal | 44 g    | 29 g   | 78 g   |
| Lose It!     | 34.2 g  | 36.5 g | 61.3 g |
| FatSecret    | 42.88 g | 46.02 g| 38.98 g|
| CRON-O-Meter | 30.1 g  | 8.1 g  | 74.1 g |

Table 17: The Protein data comparison for User 3

|              | Day 1  | Day 2  | Day 3  |
|--------------|--------|--------|--------|
| MyFitnessPal | 45 g   | 23 g   | 50 g   |
| Lose It!     | 47.6 g | 35.9 g | 53.9 g |
| FatSecret    | -      | -      | -      |
| CRON-O-Meter | 65 g   | 31.6 g | 45.1 g |

Table 16: The Sugar data comparison for User 3

|              | Day 1   | Day 2   | Day 3   |
|--------------|---------|---------|---------|
| MyFitnessPal | 36 g    | 34 g    | 23 g    |
| Lose It!     | 19.1 g  | 37 g    | 20.8 g  |
| FatSecret    | 30.54 g | 37.71 g | 19.49 g |
| CRON-O-Meter | 43.5 g  | 15.6 g  | 73.3 g  |

Table 18: The Fat data comparison for User 3

|              | Day 1  | Day 2 | Day 3  |
|--------------|--------|-------|--------|
| MyFitnessPal | 10 g   | 11 g  | 9 g    |
| Lose It!     | 11.5 g | 7.1 g | 3.8 g  |
| FatSecret    | -      | -     | -      |
| CRON-O-Meter | 16.1 g | 8.2 g | 15.8 g |

Table 19: The Fiber data comparison for User 3

*3.3.2. Overall Caloric Output*

Meal tracking is essential to promote healthy lifestyle. A study suggested that when combined with healthy diet and exercise plan, noting down what we eat can double weight loss [20]. However, getting inconsistent data can lead to further dietary related complications and mislead users to think they consumed more or less of an item, which could be biased. With all data variation given by the apps, this clearly leads to inconsistency in the overall caloric output provided for users as proof of their daily dietary consumption.

The findings revealed different caloric outputs for all subjects. Some of the differences was big, for example, the the caloric output for user #1 by MyFitnessPal and CRON-O-Meter was *1524 cal* and *1219 cal*, respectively for the same day (See Table-20). For user #2, MyFitnessPal and CRON-O-Meter provided values *839 cal* and *581 cal*, respectively (see Table-21). For user #3, MyFitnessPal and CRON-O-Meter provided values *1614 cal* and *1392 cal*, respectively (see Table-22). For user #4, MyFitnessPal and CRON-O-Meter provided values *866 cal* and *490 cal*, respectively (see Table-23). For user #5, MyFitnessPal and CRON-O-Meter provided values *1200 cal* and *1473 cal*, respectively (see Table-24). Finally, for user #6, MyFitnessPal and CRON-O-Meter provided values *1293 cal* and *1291.3* *cal*, respectively (see Table-25).

| User #1 | Day 1 | Day 2 | Day 3 |
|---|---|---|---|
| MyFitnessPal | *1524 cal* | *1255 cal* | *1480 cal* |
| Lose It! | - | - | - |
| FatSecret | *1339 cal* | *1257 cal* | *1296 cal* |
| CRON-O-Meter | *1219 cal* | *1955 cal* | *1227 cal* |

Table 20: The Amount of Caloric Output for User #1

| User #2 | Day 1 | Day 2 | Day 3 |
|---|---|---|---|
| MyFitnessPal | *839 cal* | *704 cal* | *280 cal* |
| Lose It! | - | - | - |
| FatSecret | *414 cal* | *346 cal* | *547 cal* |
| CRON-O-Meter | *581 cal* | *587.5 cal* | - |

Table 21: The Amount of Caloric Output for User #2

| User #3 | Day 1 | Day 2 | Day 3 |
|---|---|---|---|
| MyFitnessPal | *1614 cal* | *695 cal* | *509 cal* |
| Lose It! | - | - | - |
| FatSecret | *1217 cal* | *600 cal* | *699 cal* |
| CRON-O-Meter | *1392 cal* | *455.8 cal* | *866 cal* |

Table 22: The Amount of Caloric Output for User #3

| User #4 | Day 1 | Day 2 | Day 3 |
|---|---|---|---|
| MyFitnessPal | *866 cal* | *809 cal* | *1087 cal* |
| Lose It! | - | - | - |
| FatSecret | *596 cal* | *690 cal* | *1057 cal* |
| CRON-O-Meter | *490 cal* | *956.6 cal* | *1006 cal* |

Table 23: The Amount of Caloric Output for User #4

| User #5 | Day 1 | Day 2 | Day 3 |
|---|---|---|---|
| MyFitnessPal | *1200 cal* | *1500 cal* | *1457 cal* |
| Lose It! | - | - | - |
| FatSecret | *1485 cal* | *1958 cal* | *1329 cal* |
| CRON-O-Meter | *1473 cal* | *1633 cal* | *1458 cal* |

Table 24: The Amount of Caloric Output for User #5

| User #6 | Day 1 | Day 2 | Day 3 |
|---|---|---|---|
| MyFitnessPal | *1293 cal* | *2206 cal* | *949 cal* |
| Lose It! | - | - | - |
| FatSecret | *1223 cal* | *2201cal* | *809 cal* |
| CRON-O-Meter | *1291.3 cal* | *2257 cal* | *794.2 cal* |

Table 25: The Amount of Caloric Output for User #6

## 4. Data Variation

To measure the variation between the tools, we have converted the output from calories into grams to later measure the portion size for each user and the food items. The conversion revealed that among all users, some tools had closer values than others. This could be either due to similarity in their food database or food items and data calculation. However other apps provided completely different values (see below Tables).

*4.1. Data Conversion*

We followed the standard conversion methods to convert between calories, grams and kilojoules. For that, 1 kilogram = 7716.17917647 cal, or 1000 g, when converting between calories and grams (see Table-26).

*4.2. Data Variation between Tools*

After the conversion from calories to grams, we have calculated the data variation between the tools and listed them in a matrix format. The Tables below list the data variation between tools for all users per day.

| Description | Equation | Example |
|---|---|---|
| The energy in thermochemical calories E(calth) is equal to the energy in kilojoules E(kJ) times 239.0057 | E(cal) = E(kJ) 239.0057 | Convert 0.6 kilojoules to thermochemical calories. E(cal) = 0.6kJ 239.0057 = 143.4 Cal |
| The energy in 15C calories E(cal15) is equal to the energy in kilojoules E(kJ) times 238.9201 | E(cal15) = E(kJ) 238.9201 | Convert 0.6 kilojoules to 15C calories. E(cal15) = 0.6kJ 238.9201 = 143.352 cal |
| The energy in large/food calories E(Cal) is equal to the energy in kilojoules E(kJ) times 0.239 | E(Cal) = E(kJ) 0.239 | Convert 0.6 kilojoules to food calories. E(Cal) = 0.6kJ 0.239 = 0.1434 Cal |

Table 26: The Data Conversion Units and Equations [21]

| User #1 | Day 1 | Day 2 | Day 3 |
|---|---|---|---|
| *MyFitnessPal* | 19,751 g | 1,626 g | 1,918 g |
| *Lose It!* | - | - | - |
| *FatSecret* | 17,353 g | 16,290 g | 16,796 g |
| *CRON-O-Meter* | 15,798 g | 2,534 g | 15,902 g |

Table 27: Conversion from Calories to Grams for User #2

| User #2 | Day 1 | Day 2 | Day 3 |
|---|---|---|---|
| *MyFitnessPal* | 10,873 g | 9,124 g | 363 g |
| *Lose It!* | - | - | - |
| *FatSecret* | 5,365 g | 44,841 g | 7,089 g |
| *CRON-O-Meter* | 7,530 g | 76,139 g | - |

Table 28: Conversion from Calories to Grams for User #1

| User #3 | Day 1 | Day 2 | Day 3 |
|---|---|---|---|
| *MyFitnessPal* | 20,917 g | 901 g | 6,597 g |
| *Lose It!* | - | - | - |
| *FatSecret* | 15,772 g | 78 g | 9,059 g |
| *CRON-O-Meter* | 18,040 g | 59,071 g | 11,223 g |

Table 29: Conversion from Calories to Grams for User #3

| User #4 | Day 1 | Day 2 | Day 3 |
|---|---|---|---|
| *MyFitnessPal* | 11,223 g | 10,484 g | 14,087 g |
| *Lose It!* | - | - | - |
| *FatSecret* | 77,240 g | 89,422 g | 13,698 g |
| *CRON-O-Meter* | 635 g | 123,973 g | 13,038 g |

Table 30: Conversion from Calories to Grams for User #4

| User #5 | Day 1 | Day 2 | Day 3 |
|---|---|---|---|
| *MyFitnessPal* | 1,560 g | 19 g | 18,882 g |
| *Lose It!* | - | - | - |
| *FatSecret* | 1,925 g | 25,375 g | 17,224 g |
| *CRON-O-Meter* | 1,909 g | 21,163 g | 18,895 g |

Table 31: Conversion from Calories to Grams for User #5

| User #6 | Day 1 | Day 2 | Day 3 |
|---|---|---|---|
| *MyFitnessPal* | 16,757 g | 28,589 g | 12,299 g |
| *Lose It!* | - | - | - |
| *FatSecret* | 15,850 g | 28,524 g | 10,484 g |
| *CRON-O-Meter* | 16,735 g | 29,250 g | 10,293 g |

Table 32: Conversion from Calories to Grams for User #6

## 5. Comparing Nutritional Data of Tools with The Swiss FCD

After obtaining the overall nutritional data from each tool and calculating their caloric output, in this step we compare their nutritional facts output (namely, protein, sugar, fiber, fat, carbs) with the data provided by the Swiss Food Composition Database. The goal is to measure their similarity/differences with the Swiss FCD, with respect to nutritional facts and potion size.

*5.1. Data Significance & Portion Size Variation*

To calculate the similarities/differences between these tools and compare them with the Swiss FCD, we selected a sample of 13 food items that were fully available and defined by each tool. We selected food items that are rich in one of the nutritional facts we targeted in this study, namely, protein, sugar, fiber, fat, carbs, and dairy. We listed the items with their nutrients and the caloric output provided by each tool. We then searched for these items in the Swiss FCD and inserted it into the table together with the nutrients and caloric outputs given by the Swiss FCD. The measuring unit was for 1 portion size and the matrix unit was either per 100g edible portion for solid food items, or per 100ml food volume for liquid food items (see Tables-Data Sample). Although the result showed patterns of similarity between food items among the tools or tools and the Swiss FCD, however, there was still some other variations among the tools and with the Swiss FCD. There is a need to understand the reason behind this variation, whether it's the result of a calculation, or the variation coexisted even in the FCDs used by these tools. We consider this as a future work and highlight room for more research to be carried out on food journaling tools and food composition databases.

| Day 1   |     |     |       |       |
|---------|-----|-----|-------|-------|
| User #1 | MFP | LIT | FS    | COM   |
| MFP     | -   | x   | 2,398 | 3,953 |
| LIT     | -   | -   | x     | x     |
| FS      | -   | -   | -     | 1,555 |
| COM     | -   | -   | -     | -     |

Table 33: Data Variation between Tools for User #1

| Day 1   |     |     |       |       |
|---------|-----|-----|-------|-------|
| User #2 | MFP | LIT | FS    | COM   |
| MFP     | -   | x   | 5,508 | 3,343 |
| LIT     | -   | -   | x     | x     |
| FS      | -   | -   | -     | 2,165 |
| COM     | -   | -   | -     | -     |

Table 34: Data Variation between Tools for User #2

| Day 1   |     |     |       |       |
|---------|-----|-----|-------|-------|
| User #3 | MFP | LIT | FS    | COM   |
| MFP     | -   | x   | 5,145 | 2,877 |
| LIT     | -   | -   | x     | x     |
| FS      | -   | -   | -     | 2,268 |
| COM     | -   | -   | -     | -     |

Table 35: Data Variation between Tools for User #3

| Day 1   |     |     |        |     |
|---------|-----|-----|--------|-----|
| User #4 | MFP | LIT | FS     | COM |
| MFP     | -   | x   | 66,017 | 623 |
| LIT     | -   | -   | x      | x   |
| FS      | -   | -   | -      | 557 |
| COM     | -   | -   | -      | -   |

Table 36: Data Variation between Tools for User #4

| Day 1   |     |     |       |       |
|---------|-----|-----|-------|-------|
| User #5 | MFP | LIT | FS    | COM   |
| MFP     | -   | x   | 0,365 | 0,35  |
| LIT     | -   | -   | x     | x     |
| FS      | -   | -   | -     | 0,016 |
| COM     | -   | -   | -     | -     |

Table 37: Data Variation between Tools for User#5

| Day 1   |     |     |       |        |
|---------|-----|-----|-------|--------|
| User #6 | MFP | LIT | FS    | COM    |
| MFP     | -   | x   | 0,907 | 0,022  |
| LIT     | -   | -   | x     | x      |
| FS      | -   | -   | -     | 0,0885 |
| COM     | -   | -   | -     | -      |

Table 38: Data Variation between Tools for User #6

| FOOD ITEM | CARBS | PROTEIN | SUGAR | FAT | FIBER | UNIT | CALORIES | UNIT | PORTION SIZE | MATRIX UNIT | TOOLS AND DB |
|---|---|---|---|---|---|---|---|---|---|---|---|
| Lasagne di carne, preparata | 18 | 5 | 3 | 6 | 1 | Gram | 145 | Kilocalorie | 1 portion | per 100g edible portion | MYFITNESSPAL |
| | 0 | 0 | 0 | 33,2 | 0 | Gram | 478 | Kilocalorie | 1 portion | per 100g edible portion | LOSEIT |
| | 35,43 | 20,52 | 3 | 12,38 | 2 | Gram | 336 | Kilocalorie | 1 portion | per 100g edible portion | FATSECRET |
| | 22,9 | 21,1 | 3,8 | 14,9 | 2,5 | Gram | 320,89 | Kilocalorie | 1 portion | per 100g edible portion | CRONOMETER |
| | 12,9 | 9,7 | 2,7 | 6,9 | 1,6 | Gram | 15,576 | Kilocalorie | 1 portion | per 100g edible portion | SWISS FCD |

| FOOD ITEM | CARBS | PROTEIN | SUGAR | FAT | FIBER | UNIT | CALORIES | UNIT | PORTION SIZE | MATRIX UNIT | TOOLS AND DB |
|---|---|---|---|---|---|---|---|---|---|---|---|
| Lenticchie, intere, cotte (senza aggiunta di grassi e sale) | 40 | 18 | 4 | 1 | 16 | Gram | 230 | Kilocalorie | 1 portion | per 100g edible portion | MYFITNESSPAL |
| | 44.6 | 25 | 13 | 0.7 | 163 | Gram | 317 | Kilocalorie | 1 portion | per 100g edible portion | LOSEIT |
| | 36,71 | 16,44 | 3,29 | 13,25 | 14,5 | Gram | 323 | Kilocalorie | 1 portion | per 100g edible portion | FATSECRET |
| | 28,3 | 17,9 | 0,8 | 0,8 | 11,6 | Gram | 229,68 | Kilocalorie | 1 portion | per 100g edible portion | CRONOMETER |
| | 18,5 | 8,8 | 0,4 | 0,4 | 4,1 | Gram | 12,232 | Kilocalorie | 1 portion | per 100g edible portion | SWISS FCD |

| FOOD ITEM | CARBS | PROTEIN | SUGAR | FAT | FIBER | UNIT | CALORIES | UNIT | PORTION SIZE | MATRIX UNIT | TOOLS AND DB |
|---|---|---|---|---|---|---|---|---|---|---|---|
| Pizza Margherita, cotta nel forno | 53 | 6 | 13 | 6 | 4 | Gram | 271 | Kilocalorie | 1 portion | per 100g edible portion | MYFITNESSPAL |
| | 44,4 | 14 | 2,4 | 116 | 2,2 | Gram | 333 | Kilocalorie | 1 portion | per 100g edible portion | LOSEIT |
| | 34,23 | 7,13 | 0,5 | 3,9 | 1,5 | Gram | 204 | Kilocalorie | 1 portion | per 100g edible portion | FATSECRET |
| | 44,3 | 15,1 | 3,7 | 17 | 3,2 | Gram | 400,44 | Kilocalorie | 1 portion | per 100g edible portion | CRONOMETER |
| | 29,9 | 9,9 | 1,5 | 8,4 | 1,7 | Gram | 23,890 | Kilocalorie | 1 portion | per 100g edible portion | SWISS FCD |

| FOOD ITEM | CARBS | PROTEIN | SUGAR | FAT | FIBER | UNIT | CALORIES | UNIT | PORTION SIZE | MATRIX UNIT | TOOLS AND DB |
|---|---|---|---|---|---|---|---|---|---|---|---|
| Latte scremato, UHT | 11 | 8 | 11 | 0 | 0 | Gram | 80 | Kilocalorie | 1 portion | per 100ml food volume | MYFITNESSPAL |
| | 5 | 3,2 | 5 | 1,6 | 0 | Gram | 47 | Kilocalorie | 1 portion | per 100ml food volume | LOSEIT |
| | 11,88 | 8,35 | 12,47 | 0,44 | 0 | Gram | 86 | Kilocalorie | 1 portion | per 100ml food volume | FATSECRET |
| | 12,2 | 8,3 | 12,5 | 0,2 | 0 | Gram | 83,3 | Kilocalorie | 1 portion | per 100ml food volume | CRONOMETER |
| | 4,7 | 3,4 | 4,7 | 0,1 | 0 | Gram | 33,685 | Kilocalorie | 1 portion | per 100ml food volume | SWISS FCD |

| FOOD ITEM | CARBS | PROTEIN | SUGAR | FAT | FIBER | UNIT | CALORIES | UNIT | PORTION SIZE | MATRIX UNIT | TOOLS AND DB |
|---|---|---|---|---|---|---|---|---|---|---|---|
| Vitello, bistecca, alla griglia, cottura media (senza aggiunta di grassi o sale) | 0 | 6 | 0 | 2 | 0 | Gram | 41 | Kilocalorie | 1 portion | per 100g edible portion | MYFITNESSPAL |
| | 0 | 34,2 | 0 | 3,6 | 0 | Gram | 169 | Kilocalorie | 1 portion | per 100g edible portion | LOSEIT |
| | 40 | 55 | 0 | 35 | 7 | Gram | 690 | Kilocalorie | 1 portion | per 100g edible portion | FATSECRET |
| | 0 | 8,5 | 1,3 | 0 | 0 | Gram | 45,36 | Kilocalorie | 1 portion | per 100g edible portion | CRONOMETER |
| | 0 | 28,6 | 0 | 5,1 | 0 | Gram | 16,126 | Kilocalorie | 1 portion | per 100g edible portion | SWISS FCD |

| FOOD ITEM | CARBS | PROTEIN | SUGAR | FAT | FIBER | UNIT | CALORIES | UNIT | PORTION SIZE | MATRIX UNIT | TOOLS AND DB |
|---|---|---|---|---|---|---|---|---|---|---|---|
| Pesce (in media), filetto, al vapore (senza aggiunta di grassi o sale) | 1 | 20 | 0 | 3 | 0 | Gram | 100 | Kilocalorie | 1 portion | per 100g edible portion | MYFITNESSPAL |
| | 11 | 22 | 0 | 4 | 4 | Gram | 169 | Kilocalorie | 1 portion | per 100g edible portion | LOSEIT |
| | 0 | 26,33 | 0 | 1,38 | 0 | Gram | 127 | Kilocalorie | 1 portion | per 100g edible portion | FATSECRET |
| | 15,7 | 21,9 | 1 | 3,9 | 0,8 | Gram | 195,8 | Kilocalorie | 1 portion | per 100g edible portion | CRONOMETER |
| | 0 | 25,3 | 0 | 7,8 | 0 | Gram | 17,177 | Kilocalorie | 1 portion | per 100g edible portion | SWISS FCD |

| FOOD ITEM | CARBS | PROTEIN | SUGAR | FAT | FIBER | UNIT | CALORIES | UNIT | PORTION SIZE | MATRIX UNIT | TOOLS AND DB |
|---|---|---|---|---|---|---|---|---|---|---|---|
| Succo d'arancia | 26 | 2 | 21 | 0 | 0 | Gram | 112 | Kilocalorie | 1 portion | per 100ml food volume | MYFITNESSPAL |
| | 9 | 0,6 | 9 | 0,2 | 0,3 | Gram | 43 | Kilocalorie | 1 portion | per 100ml food volume | LOSEIT |
| | 25,79 | 1,74 | 20,83 | 0,5 | 0,5 | Gram | 112 | Kilocalorie | 1 portion | per 100ml food volume | FATSECRET |
| | 25,3 | 1,7 | 20,8 | 0,5 | 0,5 | Gram | 111,6 | Kilocalorie | 1 portion | per 100ml food volume | CRONOMETER |
| | 11,1 | 0,5 | 11,1 | 0,5 | 0,1 | Gram | 51,603 | Kilocalorie | 1 portion | per 100ml food volume | SWISS FCD |

| FOOD ITEM | CARBS | PROTEIN | SUGAR | FAT | FIBER | UNIT | CALORIES | UNIT | PORTION SIZE | MATRIX UNIT | TOOLS AND DB |
|---|---|---|---|---|---|---|---|---|---|---|---|
| Marmellata | 60 | 1 | 0 | 0 | 0 | Gram | 245 | Kilocalorie | 1 portion | per 100g edible portion | MYFITNESSPAL |
| | 9,1 | 0,1 | 8,8 | 0,1 | 0,3 | Gram | 37 | Kilocalorie | 1 portion | per 100g edible portion | LOSEIT |
| | 13,99 | 0,11 | 8,95 | 0,3 | 0,1 | Gram | 55 | Kilocalorie | 1 portion | per 100g edible portion | FATSECRET |
| | 13,6 | 0,1 | 9,8 | 0 | 0,2 | Gram | 55,6 | Kilocalorie | 1 portion | per 100g edible portion | CRONOMETER |
| | 58,7 | 0,5 | 58,7 | 0,5 | 2,2 | Gram | 24,846 | Kilocalorie | 1 portion | per 100g edible portion | SWISS FCD |

| FOOD ITEM | CARBS | PROTEIN | SUGAR | FAT | FIBER | UNIT | CALORIES | UNIT | PORTION SIZE | MATRIX UNIT | TOOLS AND DB |
|---|---|---|---|---|---|---|---|---|---|---|---|
| Pasta, cotta in acqua salata (sale non iodato) | 8 | 2 | 2 | 1 | 2 | Gram | 45 | Kilocalorie | 1 portion | per 100g edible portion | MYFITNESSPAL |
| | - | - | - | - | - | Gram | 137 | Kilocalorie | 1 portion | per 100g edible portion | LOSEIT |
| | 14,21 | 2,94 | - | 0,6 | - | Gram | 75 | Kilocalorie | 1 portion | per 100g edible portion | FATSECRET |
| | 39,5 | 9 | 1,1 | 2,6 | 5,9 | Gram | 224,99 | Kilocalorie | 1 portion | per 100g edible portion | CRONOMETER |
| | 32,8 | 5,8 | 0,2 | 0,6 | 2,4 | Gram | 16,675 | Kilocalorie | 1 portion | per 100g edible portion | SWISS FCD |

| FOOD ITEM | CARBS | PROTEIN | SUGAR | FAT | FIBER | UNIT | CALORIES | UNIT | PORTION SIZE | MATRIX UNIT | TOOLS AND DB |
|---|---|---|---|---|---|---|---|---|---|---|---|
| Broccoli, al vapore (senza aggiunta di sale) | 6 | 3 | 2 | 0 | 2 | Gram | 31 | Kilocalorie | 1 portion | per 100g edible portion | MYFITNESSPAL |
| | 7,8 | 4,4 | 3,3 | 0,6 | 4,4 | Gram | 45 | Kilocalorie | 1 portion | per 100g edible portion | LOSEIT |
| | 6,04 | 2,57 | 1,55 | 0,34 | 2,4 | Gram | 31 | Kilocalorie | 1 portion | per 100g edible portion | FATSECRET |
| | 3,7 | 2,6 | 1,5 | 0,3 | 2,4 | Gram | 3094 | Kilocalorie | 1 portion | per 100g edible portion | CRONOMETER |
| | 2,3 | 2,9 | 2,2 | 0,4 | 2,9 | Gram | 30,102 | Kilocalorie | 1 portion | per 100g edible portion | SWISS FCD |

| FOOD ITEM | CARBS | PROTEIN | SUGAR | FAT | FIBER | UNIT | CALORIES | UNIT | PORTION SIZE | MATRIX UNIT | TOOLS AND DB |
|---|---|---|---|---|---|---|---|---|---|---|---|
| Cotoletta (media di vitello, maiale e agnello), alla griglia (senza aggiunta di grassi o sale) | 13 | 12 | 0 | 14 | 0 | Gram | 229 | Kilocalorie | 1 portion | per 100g edible portion | MYFITNESSPAL |
| | - | - | - | - | - | Gram | 350 | Kilocalorie | 1 portion | per 100g edible portion | LOSEIT |
| | 8,38 | 23,21 | 0 | 7,81 | 0,3 | Gram | 194 | Kilocalorie | 1 portion | per 100g edible portion | FATSECRET |
| | 0 | 70,7 | 0 | 24,9 | 0 | Gram | 525,24 | Kilocalorie | 1 portion | per 100g edible portion | CRONOMETER |
| | 0 | 30 | 0 | 14,4 | 0 | Gram | 24,846 | Kilocalorie | 1 portion | per 100g edible portion | SWISS FCD |

| FOOD ITEM | CARBS | PROTEIN | SUGAR | FAT | FIBER | UNIT | CALORIES | UNIT | PORTION SIZE | MATRIX UNIT | TOOLS AND DB |
|---|---|---|---|---|---|---|---|---|---|---|---|
| Insalata belga, cruda | 0 | 0 | 0 | 0 | 0 | Gram | 14 | Kilocalorie | 1 portion | per 100g edible portion | MYFITNESSPAL |
| | 4,3 | 1,4 | 0,8 | 0,3 | 3,4 | Gram | 19 | Kilocalorie | 1 portion | per 100g edible portion | LOSEIT |
| | 2,32 | 0,55 | 1,33 | 0,11 | 0,8 | Gram | 11 | Kilocalorie | 1 portion | per 100g edible portion | FATSECRET |
| | 10,6 | 1 | 9,4 | 0,2 | 1,5 | Gram | 55,27 | Kilocalorie | 1 portion | per 100g edible portion | CRONOMETER |
| | 0,7 | 1 | 0,6 | 0,2 | 2,5 | Gram | 13,379 | Kilocalorie | 1 portion | per 100g edible portion | SWISS FCD |

| FOOD ITEM | CARBS | PROTEIN | SUGAR | FAT | FIBER | UNIT | CALORIES | UNIT | PORTION SIZE | MATRIX UNIT | TOOLS AND DB |
|---|---|---|---|---|---|---|---|---|---|---|---|
| Pomodoro, crudo | 0 | 0 | 0 | 0 | 0 | Gram | 15 | Kilocalorie | 1 portion | per 100g edible portion | MYFITNESSPAL |
| | 7,6 | 1,4 | 5,8 | 0,5 | 2,2 | Gram | 38 | Kilocalorie | 1 portion | per 100g edible portion | LOSEIT |
| | 4,82 | 1,08 | 3,23 | 0,25 | 1,5 | Gram | 22 | Kilocalorie | 1 portion | per 100g edible portion | FATSECRET |
| | 2,5 | 1,3 | 2,9 | 0,2 | 1 | Gram | 17,8 | Kilocalorie | 1 portion | per 100g edible portion | CRONOMETER |
| | 3,2 | 0,8 | 3,2 | 0,3 | 1,2 | Gram | 21,262 | Kilocalorie | 1 portion | per 100g edible portion | SWISS FCD |

## 5.2. Data Significance

After listing the 13 food items, we checked for the significances of their data in terms of grams. For that, we converted the caloric output for each tool and the Swiss DB into grams. We followed a standard conversion equation, that's 1 gram contains 7.71617917647 calories, and 1 kilogram is equal to 7716.17917647 calories, or 1000 grams. We list our result in the below Tables representing the caloric output in grams for the 13 food items selected. We represented the data in matrix form.

Table 39: Amount of Caloric Data Significant for Sample 1

| Lasagne di carne, preparata | | | | | |
|---|---|---|---|---|---|
| | MyFitnessPal | Lose It! | FatSecret | CRON-O-Meter | Swiss FCD |
| MyFitnessPal | x | 43 | 25 | 23 | 17 |
| Lose It! | 0 | x | 18 | 16 | 60 |
| FatSecret | 0 | 0 | x | 2 | 42 |
| CRON-O-Meter | 0 | 0 | 0 | x | 40 |
| Swiss FCD | 0 | 0 | 0 | 0 | x |

Table 40: Amount of Caloric Data Significant for Sample 2

| Lenticchie, intere, cotte (senza aggiunta di grassi e sale) | | | | | |
|---|---|---|---|---|---|
|  | *MyFitnessPal* | *Lose It!* | *FatSecret* | *CRON-O-Meter* | *Swiss FCD* |
| *MyFitnessPal* | x | 11 | 12 | 0 | 28 |
| *Lose It!* | 0 | x | 1 | 11 | 39 |
| *FatSecret* | 0 | 0 | x | 12 | 40 |
| *CRON-O-Meter* | 0 | 0 | 0 | x | 28 |
| *Swiss FCD* | 0 | 0 | 0 | 0 | x |

Table 41: Amount of Caloric Data Significant for Sample 3

| Pizza Margherita, cotta nel forno | | | | | |
|---|---|---|---|---|---|
|  | *MyFitnessPal* | *Lose It!* | *FatSecret* | *CRON-O-Meter* | *Swiss FCD* |
| *MyFitnessPal* | x | 12 | 11 | 23 | 40 |
| *Lose It!* | 0 | x | 23 | 11 | 40 |
| *FatSecret* | 0 | 0 | x | 34 | 23 |
| *CRON-O-Meter* | 0 | 0 | 0 | x | 49 |
| *Swiss FCD* | 0 | 0 | 0 | 0 | x |

Table 42: Amount of Caloric Data Significant for Sample 4

| Latte scremato, UHT | | | | | |
|---|---|---|---|---|---|
|  | *MyFitnessPal* | *Lose It!* | *FatSecret* | *CRON-O-Meter* | *Swiss FCD* |
| *MyFitnessPal* | x | 4 | 1 | 1 | 6 |
| *Lose It!* | 0 | x | 5 | 5 | 2 |
| *FatSecret* | 0 | 0 | x | 0 | 7 |
| *CRON-O-Meter* | 0 | 0 | 0 | x | 7 |
| *Swiss FCD* | 0 | 0 | 0 | 0 | x |

Table 43: Amount of Caloric Data Significant for Sample 5

| Vitello, bistecca, alla griglia, cottura media (senza aggiunta di grassi o sale) | | | | | |
|---|---|---|---|---|---|
|  | *MyFitnessPal* | *Lose It!* | *FatSecret* | *CRON-O-Meter* | *Swiss FCD* |
| *MyFitnessPal* | x | 17 | 84 | 1 | 3 |
| *Lose It!* | 0 | x | 67 | 16 | 20 |
| *FatSecret* | 0 | 0 | x | 83 | 87 |
| *CRON-O-Meter* | 0 | 0 | 0 | x | 4 |
| *Swiss FCD* | 0 | 0 | 0 | 0 | x |

Table 44: Amount of Caloric Data Significant for Sample 6

| Pesce (in media), filetto, al vapore (senza aggiunta di grassi o sale) | | | | | |
|---|---|---|---|---|---|
| | *MyFitnessPal* | *Lose It!* | *FatSecret* | *CRON-O-Meter* | *Swiss FCD* |
| *MyFitnessPal* | x | 10 | 4 | 13 | 10 |
| *Lose It!* | 0 | x | 6 | 3 | 20 |
| *FatSecret* | 0 | 0 | x | 9 | 14 |
| *CRON-O-Meter* | 0 | 0 | 0 | x | 23 |
| *Swiss FCD* | 0 | 0 | 0 | 0 | x |

Table 45: Amount of Caloric Data Significant for Sample 7

| Succo d'arancia | | | | | |
|---|---|---|---|---|---|
| | *MyFitnessPal* | *Lose It!* | *FatSecret* | *CRON-O-Meter* | *Swiss FCD* |
| *MyFitnessPal* | x | 9 | 1 | 1 | 8 |
| *Lose It!* | 0 | x | 8 | 8 | 1 |
| *FatSecret* | 0 | 0 | x | 0 | 7 |
| *CRON-O-Meter* | 0 | 0 | 0 | x | 7 |
| *Swiss FCD* | 0 | 0 | 0 | 0 | x |

Table 46: Amount of Caloric Data Significant for Sample 8

| Marmellata | | | | | |
|---|---|---|---|---|---|
| | *MyFitnessPal* | *Lose It!* | *FatSecret* | *CRON-O-Meter* | *Swiss FCD* |
| *MyFitnessPal* | x | 27 | 25 | 25 | 29 |
| *Lose It!* | 0 | x | 2 | 2 | 2 |
| *FatSecret* | 0 | 0 | x | 0 | 4 |
| *CRON-O-Meter* | 0 | 0 | 0 | x | 4 |
| *Swiss FCD* | 0 | 0 | 0 | 0 | x |

Table 47: Amount of Caloric Data Significant for Sample 9

| Pasta, cotta in acqua salata (sale non iodato) | | | | | |
|---|---|---|---|---|---|
| | *MyFitnessPal* | *Lose It!* | *FatSecret* | *CRON-O-Meter* | *Swiss FCD* |
| *MyFitnessPal* | x | 12 | 4 | 23 | 4 |
| *Lose It!* | 0 | x | 8 | 11 | 16 |
| *FatSecret* | 0 | 0 | x | 19 | 8 |
| *CRON-O-Meter* | 0 | 0 | 0 | x | 27 |
| *Swiss FCD* | 0 | 0 | 0 | 0 | x |

Table 48: Amount of Caloric Data Significant for Sample 10

| Broccoli, al vapore (senza aggiunta di sale) | | | | | |
|---|---|---|---|---|---|
| | MyFitnessPal | Lose It! | FatSecret | CRON-O-Meter | Swiss FCD |
| MyFitnessPal | x | 2 | 0 | 36 | 0 |
| Lose It! | 0 | x | 2 | 34 | 2 |
| FatSecret | 0 | 0 | x | 36 | 0 |
| CRON-O-Meter | 0 | 0 | 0 | x | 36 |
| Swiss FCD | 0 | 0 | 0 | 0 | x |

Table 49: Amount of Caloric Data Significant for Sample 11

| Cotoletta (media di vitello, maiale e agnello), alla griglia (senza aggiunta di grassi o sale) | | | | | |
|---|---|---|---|---|---|
| | MyFitnessPal | Lose It! | FatSecret | CRON-O-Meter | Swiss FCD |
| MyFitnessPal | x | 15 | 5 | 38 | 27 |
| Lose It! | 0 | x | 20 | 23 | 42 |
| FatSecret | 0 | 0 | x | 43 | 22 |
| CRON-O-Meter | 0 | 0 | 0 | x | 65 |
| Swiss FCD | 0 | 0 | 0 | 0 | x |

Table 50: Amount of Caloric Data Significant for Sample 12

| Insalata belga, cruda | | | | | |
|---|---|---|---|---|---|
| | MyFitnessPal | Lose It! | FatSecret | CRON-O-Meter | Swiss FCD |
| MyFitnessPal | x | 0 | 1 | 5 | 0 |
| Lose It! | 0 | x | 1 | 5 | 0 |
| FatSecret | 0 | 0 | x | 5 | 1 |
| CRON-O-Meter | 0 | 0 | 0 | x | 5 |
| Swiss FCD | 0 | 0 | 0 | 0 | x |

Table 51: Amount of Caloric Data Significant for Sample 13

| Pomodoro, crudo | | | | | |
|---|---|---|---|---|---|
| | MyFitnessPal | Lose It! | FatSecret | CRON-O-Meter | Swiss FCD |
| MyFitnessPal | x | 3 | 1 | 0 | 1 |
| Lose It! | 0 | x | 2 | 3 | 2 |
| FatSecret | 0 | 0 | x | 1 | 0 |
| CRON-O-Meter | 0 | 0 | 0 | x | 1 |
| Swiss FCD | 0 | 0 | 0 | 0 | x |

## 6. Findings

Food journaling is a way to evaluate eating habits. Combined with goal setting and group therapy sessions, food journaling emphasizes the importance of monitoring and modifying daily behaviour. In addition, diet and food tracking have an impact on weight loss and chronic disease prevention. Dietary related behaviours have an effect on user's lifestyle and overall health condition. However, the way existing food journaling tools track diet isn't sufficient, the data provided by these tools aren't validated and consistent. Getting different feedbacks among the apps can provide wrong information and lead to complications. The conducted experimentation with 6 users (5 male, 1 female) showed differences in the data output by each tool. Moreover, some food items were missing from some apps. Although only 4 apps used in the measure, which is a small subset of the available tools, the obtained result revealed some inconsistent and missing data among these tools. Moreover, the data obtained from the tools and compared with the Swiss FCD showed some inconsistency, although the similarity in terms of values among some tools. The overall caloric value outputted varies greatly among the tools and the food database. This questions the reliability and effectiveness of these tools, especially when intended for diet and weight loss tracking.

## 7. Discussion

The usability and applicability of food journaling application for diet and weight loss management have been largely investigated [22]. However, to our knowledge, this study is the first to compare output of food journaling tools and measure their consistency and compatibility. We performed in-depth exploration on the output data of such tools and the correlation between the data for the same food item. The data collection lasted for 10 days and we had a small number of users, this however was sufficient to validate our hypothesis. Participants used their own smartphone, which allowed them to be familiar with the application. Our results are applicable outside the research setting and to measure the efficacy of similar apps in different domains. Future direction should consider discussing reasons behind the variation in the output among these tools and create a consistent data where all tools share the same value for the same item. We should investigate why this variation occurs and if there is a correlation between the food database used and the unit calculation applied. In addition, there is a need to investigate where the error occurs, whether its on user side while food logging and estimating portion, on the app side while calculating or its generated by the food database.

## 8. Conclusion

With the advancement of technology, methods for monitoring food logging and other food related parameters have become simpler and more convenient. This paper characterized food journaling tools to track user diet for 10 days. The aim was to measure if tools provide consistent output. the study also focused on comparing this output with the Swiss FCD. We explored interesting relationships among dietary related data provided by these food journaling apps. The results indicate some similarity between the values among some tools, but the data was inconsistent and there was a variation in the output feedback. In addition, there was another inconsistency arose when we compared these data with the data in Swiss FCD. Some tools missed essential nutritional fact data, such as sugar and fiber. Based on the evaluation of the tools, there is a gap in the data they provided, and this questions their effectiveness to follow for food logging and diet tracking. Research is needed to explore ways to provide people with a self-monitoring method fitting their preferences. Overall, we pointed to some data related issues of electronic methods for self-monitoring of diet. Finally, ways to predict which self-monitoring method works best for an individual are needed. We envisage this study to contribute to the current research in mobile apps for mHealth and fitness and be used as a reference to understand the current positions for improvements in this domain.

## Acknowledgement

We thank all the students who dedicated their time and effort to participate in this experiment and make it happen.